\documentclass[aps,prb,amsmath,amssymb,twocolumn]{revtex4}
\usepackage{graphicx}
\usepackage{hyperref}

\begin{document}

\title{Edge states in two-dimensional electron gas with heterogeneous spin-orbit interaction}

\author{Aleksei~A.~Sukhanov and Vladimir~A.~Sablikov}

\affiliation{V.A.~Kotel'nikov Institute of Radio Engineering and Electronics,
Russian Academy of Sciences, Fryazino, Moscow District, 141190,
Russia}

\begin{abstract}
We show that edge states similar to those known for topological insulators exist in two-dimensional electron system with one-band spectrum in the presence of heterogeneous spin-orbit interaction (SOI). These states appear at boundaries between regions with the SOIs of different kind or between the regions with the SOI and without it. Depending on the system parameters they can appear in an energy range lying both in the forbidden and conduction bands of bulk states. The edge states have chiral spin texture and carry a spin current under the equilibrium. We study also the size quantization of the edge states in a strip structure with two boundaries to find an unusual dependence of the quantization energy on the strip width.
\end{abstract}
    
\maketitle
Spin-orbit interaction (SOI) in solids generates a wealth of fascinating effects which are somehow connected with the presence of boundaries and interfaces. It is enough to mention such effects as: the quantum spin Hall effect,~\cite{Dyakonov,Murakami} spin accumulation in the regime of extrinsic spin Hall effect,~\cite{Kato} spin-dependent tunneling,~\cite{Perel,Sablikov} edge spin currents,~\cite{Sablikov1} edge spin accumulation,~\cite{Sonin} etc.

Presently increasing attention is paid to topological insulators~\cite{Hasan,Zhang,Moore} which genetically originates from the SOIs. Topological insulators are characterized by the presence of edge or surface  states (correspondingly in two- or three-dimensional cases) lying in the energy gap of bulk states. An important property of topological states is their chiral spin texture correlated with the wave vector, due to which these states are topologically protected against the scattering and carry a dissipationless spin current. The dispersion curve of these states connects the valence and conduction bands so the simplest model describing them uses a two-band Hamiltonian.~\cite{Volkov,Suris,Korenman}

In this paper we show that somewhat similar states exist also in a much more simple system where they appear even within one-band model. We find these states in two-dimensional (2D) electron systems containing a boundary between regions with the SOIs of different kind, such as a contact of regions with the Rashba and Dresselhaus SOIs. In particular, the edge states exist in a contact of 2D regions with the SOI and the normal 2D electron gas without SOI. The distinctive feature of these states is that the dispersion curve of their spectrum goes from the conduction band, circumscribes a loop in the forbidden band and then turns to the conduction band.

Below we study the spectra of the edge states, the spin density distribution and spin-currents in structures RSOI/N and RSOI/DSOI, where N denotes normal 2D electron gas, RSOI and DSOI stand for regions with the Rashba and Dresselhaus SOIs. In addition, we consider the size quantization effect on the edge states in strip structures DSOI/RSOI/DSOI where the central RSOI region is of finite width. 

The 2D system with heterogeneous SOI is described by the Hamiltonians 
\begin{equation}
H_i=\frac{p^{2}}{2m_i}+H_{R,D}^{(i)}+U_i\,,
\label{H}
\end{equation}
defined in each region indexed by $i$ with uniform potential $U_i$ and SOI strength. Here $m_i$ is the effective mass of electrons, $p=(p_x,p_y)$ is the momentum, $H_{R,D}^{(i)}$ is the Rashba or Dresselhaus SOI Hamiltonian, 
\begin{equation}
 H_{R}\!=\!\frac{\alpha_i}{\hbar}(p_{y}\sigma_{x}\!-\!p_{x}\sigma_{y}),\quad H_{D}\!=\!\frac{\beta_i}{\hbar}(p_{y}\sigma_{y}\!-\!p_{x}\sigma_{x})\,,
\end{equation}
$\sigma_x$, $\sigma_y$ are the Pauli matrices. We use the effective mass approximation which is well justified since the envelope wave functions are supposed to vary slowly at the lattice constant scale.~\cite{EMA}

To be specific, we consider the case where $x$ axis is normal to the boundary and $y$ axis is parallel to it. The wave functions $\mbox{\boldmath$\Psi$}^{(i)}$ are determined by the Schr\"odinger equation with boundary conditions~\cite{Sablikov}
\begin{equation}
 \mbox{\boldmath$\Psi$}^{(i)} \Big|_{-0}^{+0}=0\,, \quad
 \left[\frac{1}{m_i}\frac{\partial}{\partial x}-\mathcal{B}_i^{(R,D)}\right]_{-0}^{+0} \mbox{\boldmath$\Psi$}^{(i)}=0\,,
\label{bound_cond}
\end{equation}
where $\mathcal{B}^{(R)}\!=\!i\alpha_i\sigma_y/\hbar^2$ in the case of Rashba SOI and $\mathcal{B}^{(D)}\!=\!i\beta_i\sigma_x/\hbar^2$ for Dresselhaus SOI.

General solution in $i$-th region reads as
\begin{equation}
 \mbox{\boldmath$\Psi$}^{(i)}_{k^{(i)}_x,k_y}=e^{ik_yy}\sum\limits_j^{1,2}\sum\limits_s^{\pm}A^{(i)}_{j,s} \mbox{\boldmath$\chi$}^{(i)}_{j,s}e^{ik^{(i)}_{j,s}x}\,,
\label{wave_func}
\end{equation} 
where $k_y$ is the tangential wave vector, $k^{(i)}_{j,s}$ is the $x$ component of the wave vector, defined by a characteristic equation of the Hamiltonian~(\ref{H}), $s$ stands for the spin index, $j$ numbers the solutions of the characteristic equation, $\mbox{\boldmath$\chi$}^{(i)}_{j,s}$ is the spin function. 

Generally there is a set of four wave vectors $k^{(i)}_{j,s}$, which are described in detail in Ref.~\onlinecite{Sablikov}. A short resume is as follows. The wave vectors $k^{(i)}_{j,s}$ are complex functions of the energy $E$ and the tangential momentum $k_y$. Two of them correspond to the states propagating or decreasing to the right along the $x$ axis, other two relate to the states propagating or decreasing in opposite direction. In the energy region $E<U_i-E_{so}$, all $k^{(i)}_{j,s}$ contain both real and imaginary parts which describe decaying and oscillating states. $E_{so}$ is the characteristic energy of the SOI: $E_{so}=m\alpha^2/(2\hbar^2)$ for the Rashba SOI and $E_{so}=m\beta^2/(2\hbar^2)$ for the Dresselhaus SOI. When $E>U_i-E_{so}$, the wave vectors $k^{(i)}_{j,s}$ are either purely real or imaginary depending on the relation between $E$ and $k_y$. The full analysis of the $k^{(i)}_{j,s}$ dependence on $E$ and $k_y$ is available in Refs~\onlinecite{Sablikov,Tkach,Sukhanov}.

To clarify if the edge states exist near the interface $x=0$ one needs to find the solutions satisfying the boundary conditions [Eq.~(\ref{bound_cond})] and vanishing at the infinity ($x\to\pm\infty$). Dropping the terms, which do not vanish at the infinity, in Eq.~(\ref{wave_func}) we obtain a system of homogeneous equations. The zeros of its determinant give equations for the edge state spectrum.

Below we present results of the edge-state spectrum calculations for several structures: contacts SOI/N, RSOI/DSOI and strip structure DSOI/RSOI/DSOI.   

\begin{figure}
\includegraphics[width=0.8\linewidth]{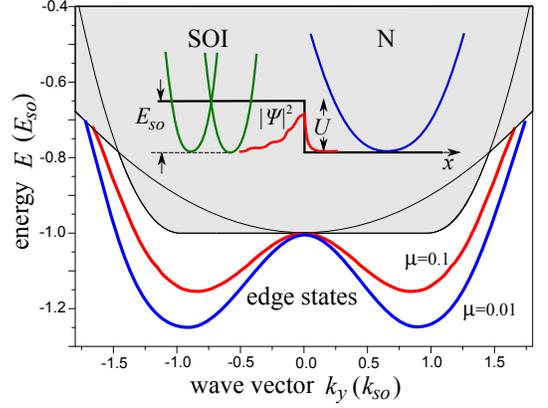}
\caption{(Color online). Spectrum of the edge states at SOI/N boundary for $m_{so}/m_N=0.1$ and $m_{so}/m_N=0.01$. Thin lines depict the boundaries of the bulk state continuum (shaded) in the SOI and N regions for $U=-E_{so}$. The inset shows schematically the potential shape, the bulk state spectra and the distribution of the electron density in the edge state.} 
\label{SOI_N_mu0}
\end{figure}

\begin{figure}
\includegraphics[width=0.8\linewidth]{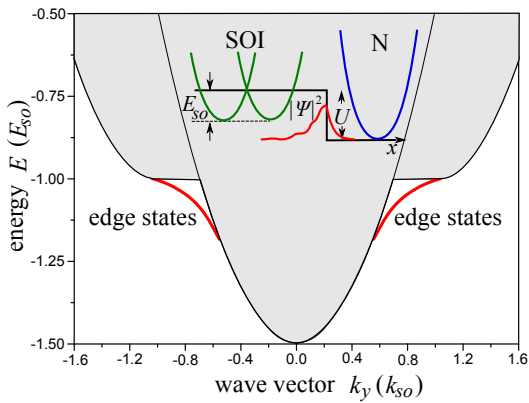}
\caption{(Color online). Spectrum of the edge states at the SOI-N boundary for $m_{so}/m_N=1$, $U=1.5E_{so}$.} 
\label{SOI_N_mu1}
\end{figure}

To begin, consider the SOI/N contact. The region with the Rashba SOI is located at $x<0$ and N region lies at $x>0$. The energy diagram is depicted in the insets in Figs~\ref{SOI_N_mu0},~\ref{SOI_N_mu1}. The potential $U_N$ in N region is lower than that in the SOI region: $U_N=-U$, $U_{so}=0$. 

Of most interest is the energy interval $E<-E_{so}$, where the electron states in the SOI region are evanescent~\cite{Sablikov} and the wave functions are
\begin{equation}
\mbox{\boldmath$\Psi$}^{(SOI)}=e^{ik_yy}e^{\kappa x}\sum \limits_{s=\pm}A_{s}\binom{\chi_{s}}{1} e^{is kx}\,,
\label{eq2}
\end{equation}
where 
\begin{align}
k&=\frac{1}{\sqrt{2}}\sqrt{\zeta\!-\!k_y^2+2k_{so}^{2}+\sqrt{\left(\zeta\!-\!k_y^2\right)^2\!-\!4k_{so}^2k_y^2}}\,,\\
\kappa&=\frac{1}{\sqrt{2}}\sqrt{-\zeta\!+\!k_y^2-2k_{so}^{2}+\sqrt{\left(\zeta\!-\!k_y^2\right)^2\!-\!4k_{so}^2k_y^2}}\,,
\label{krx}
\end{align}
\begin{equation}
 \chi_s=-k_{so}\frac{k_y+isk-\kappa}{k_{so}^2+isk\kappa}\,,
\end{equation}
$\zeta=2m_{so}E/\hbar^2$, $k_{so}=m_{so}\alpha/\hbar^2$, $m_{so}$ is the effective mass of electrons in the SOI region.

In the N region the wave function is
\begin{equation}
\mbox{\boldmath$\Psi$}^{(N)}=e^{ik_yy}\left[t_{1}\binom{1}{0}+t_{2}\binom{0}{1}\right] e^{-g x}\,,
\label{wave-n}
\end{equation}
where $g=\sqrt{k_{y}^{2}-(\zeta+u)/\mu}$, $u=2 m_{so}U/\hbar^2$, $\mu=m_{so}/m_N$, and $m_N$ is the effective mass in the N region.

By matching the wave functions at $x=0$ we come to the following condition under which the wave function amplitudes are nonzero:
\begin{equation}
 (\kappa+\mu g)^2+k^2-k_{so}^2=0\,.
\label{eq11}
\end{equation}
Taking into account that $k$, $\kappa$ and $g$ are functions of $\zeta$ and $k_y$, Eq.~(\ref{eq11}) gives the spectrum of the edge states: $\zeta=\zeta_{ES}(k_y)$. This equation is rather cumbersome in the explicit form therefore we consider two specific cases.

In the limiting case $\mu=m_{so}/m_N\to 0$, the edge state spectrum has a simple form
\begin{equation}
 \zeta_{ES}(k_y)\simeq k_y^2-|k_{so}|\sqrt{k_{so}^2+4k_y^2}\,.
\label{mu=0}
\end{equation} 
It is seen that the edge states exist below the conduction band bottom in the SOI region occupying the interval $-1.25k_{so}^2\leq \zeta \leq -k_{so}^2$. At a given energy there are two pairs of the edge states with different signs of the wave vector $k_y$ and the group velocity. If the potential of the N region is higher than $-1.25E_{so}$, the edge states are located in the forbidden band where the bulk states are absent. When the effective mass ratio $\mu$ is finite, the edge state spectrum deviates from the limiting expression~(\ref{mu=0}) as it is shown in Fig.~\ref{SOI_N_mu0} for $U_N=-E_{so}$. 

At the energy higher than the conduction band bottom in the SOI region, $E>-E_{so}$, the edge states exist against the background of the bulk states. We have studied these states by making specific calculations for this energy region. Two edge states with $k_y<0$ and $k_y>0$ are found at a given energy in the interval $-E_{so}<E<E_{cr}$, with $E_{cr}$ being the energy at which the edge-state spectrum intersects the bulk-state spectrum boundary.

When $m_{so}=m_n$, the edge state spectrum has the form
\begin{equation}
\zeta_{ES}=\left(\frac{u-k_{so}^2}{2k_{so}}\right)^2-\frac{k_{so}^{2}k_y^2}{k_{y}^{2}-[(u-k_{so}^2)/2k_{so}]^2}\,.
\label{SOI/N_mu1}
\end{equation}
The validity of Eq.~(\ref{SOI/N_mu1}) is restricted by the inequality $k_y^2\geq\zeta+u$ appearing from the requirement that $g$ to be real. The edge states are absent when this condition is violated. The edge-state spectrum has two branches corresponding to waves propagating in opposed directions. They are shown in Fig.~\ref{SOI_N_mu1} for the potential step $U=1.5E_{so}$ at the interface. The edge states occupy a finite energy layer and a finite interval of $k_y$. The lower and upper edges of this interval are determined by the intersection points of the edge-state spectrum with the boundaries of the bulk spectra in the N and SOI regions. No edge states exist below the conduction band bottom of the N region.

The edge-state band bottom $E_b$ depends on the potential step $U$ at the interface. The function $E_b(U)$ is easy to find from Eq.~(\ref{eq11}) and the condition $g(\zeta,k_y)=0$,
\begin{equation}
 E_b=-U+\frac{U^2-E_{so}^2}{4E_{so}}.
\end{equation} 
It is seen that $E_b=-E_{so}$ at $U=E_{so}$. With increasing $U$, the edge-state bottom $E_b$ decreases to reach the minimum $E_b=-1.25E_{so}$ and whereupon $E_b$ goes up. Thus, the maximum depth of the edge-state bottom is $-0.25E_{so}$ below the conduction band bottom of the SOI region. This conclusion is easy generalized to the arbitrary mass ratio $\mu$.

The edge state formation can be interpreted as being a result of the lowering of the electron energy near the interface because of the mutual penetration of electrons from one contacting region to another. Electrons  penetrating from the N region into the SOI region gain the energy since they undergo the SOI action. On the contrary, the electrons of the SOI region lose the energy while penetrating into the N region since they do not feel the SOI there. If $m_{so}\ll m_N$, the electrons penetrate into the SOI region much deeper than into the N region. Hence, the gain in the energy is larger than its loss and a state localized near the interface can appear with energy lower than the conduction band bottom.

\begin{figure}[h,t]
\includegraphics[width=0.8\linewidth]{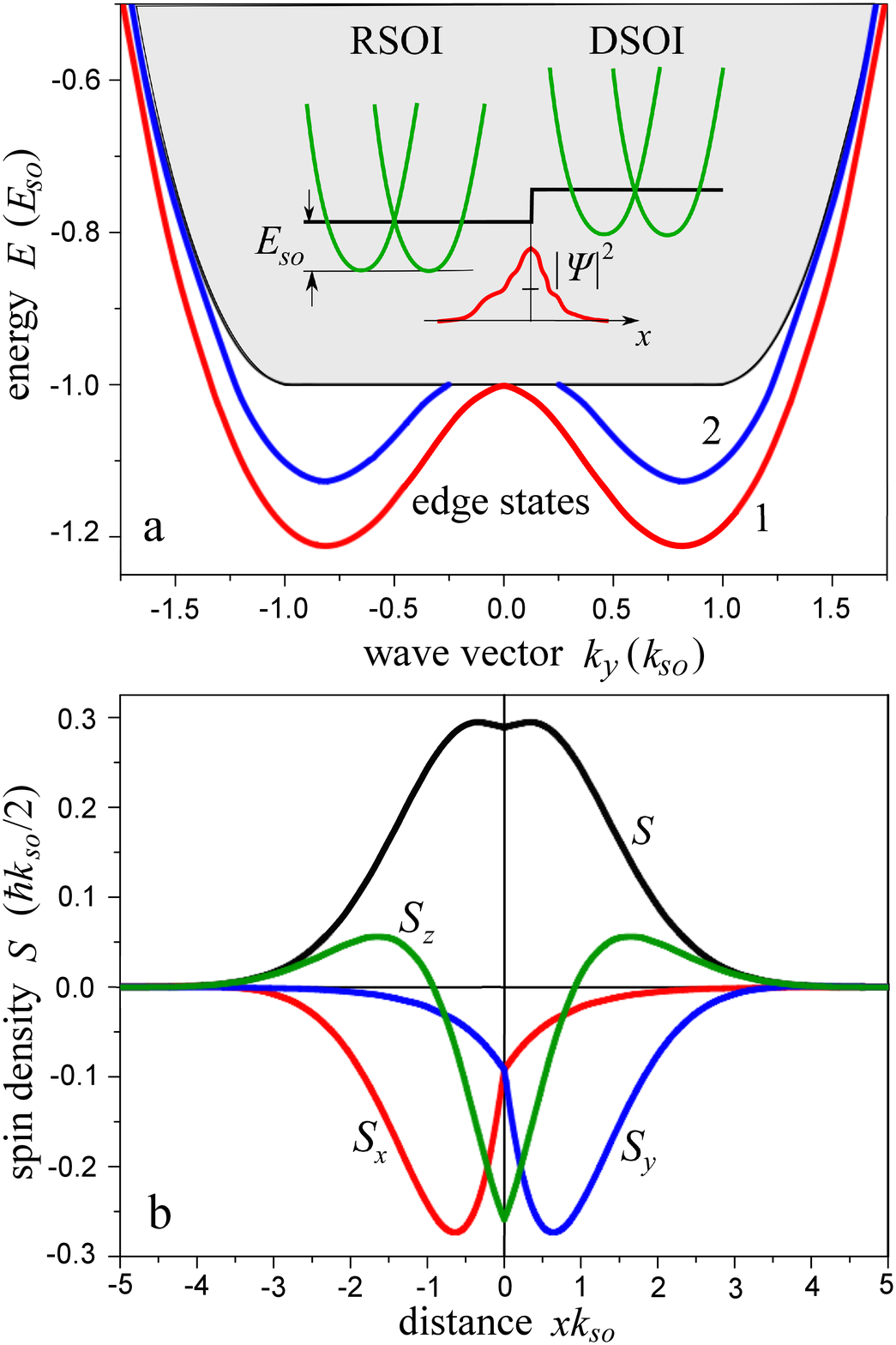}
\caption{(Color online). a) Edge-state spectrum in RSOI/DSOI structure for $U_D-U_R=0$ (line 1) and $U_D-U_R=0.25E_{so}$ (line 2). Shaded regions are the bulk states. The inset shows the potential shape, the bulk state spectra and the distribution of the electron density in the edge state. b) Spacial distribution of the spin density components ($S_x$, $S_y$, $S_z$) and the total spin density $S$ for the edge state with the energy $\zeta_{ES}=-1.2k_{so}^2$ and momentum $k_y=0.95k_{so}$ near the RSOI/DSOI boundary in the case where $U_D=U_R$, $\alpha=\beta$, $m_R=m_D$.}
\label{RSOI_DSOI}
\end{figure}

Another 2D system, in which we demonstrate the edge states in the forbidden band, is the RSOI/DSOI structure. The edge states are studied by solving Eqs.~(\ref{bound_cond}) and (\ref{wave_func}) in the same manner as described above. In the calculations we have used the expressions for wave vectors $k^{(R)}_{j,s}$ and $k^{(D)}_{j,s}$ in the Rashba and Dresselhaus regions derived in Ref.~\onlinecite{Sablikov}. The results are presented in Fig.~\ref{RSOI_DSOI} for two potential steps at the boundary.

The edge states are seen to exist in the forbidden band even if the effective masses in the contacting regions are equal, in contrast to the case of the SOI/N system. But the energy interval, where the edge states are located, and the general view of the spectra are quite similar to those of the SOI/N structure shown in Fig.~\ref{SOI_N_mu0}. The origin of the edge states can be interpreted a result of the joint action of the SOIs in both contacting regions.

The electron and spin densities in the edge states are localized near the boundary at a distance of the order of $k_{so}^{-1}$. It is worth noting that the $S_x$ component of the spin density is concentrated in the RSOI region whereas $S_y$ component is located mainly in the DSOI region (Fig.~\ref{RSOI_DSOI}b). The spin direction is reversed with changing the sign of $k_y$. Therefore the edge states are chiral ones. They are protected against the backscattering by the time-reversal symmetry. In addition, the edge states carry a spin current under the thermal equilibrium.~\cite{note1} Under the nonequilibrium conditions when a particle current flows along the boundary, the spin density accumulates in the edge states. 

Interesting effects arise when the edge states appear under the condition of size quantization. We have studed these effects by considering the strip structures DSOI/RSOI/DSOI and N/RSOI/N. In this case the calculations are more cumbersome since the wave functions should be found in three regions using two boundary conditions of the form of Eq.~\ref{bound_cond}. We have solved this problem numerically.

\begin{figure}
\includegraphics[width=0.8\linewidth]{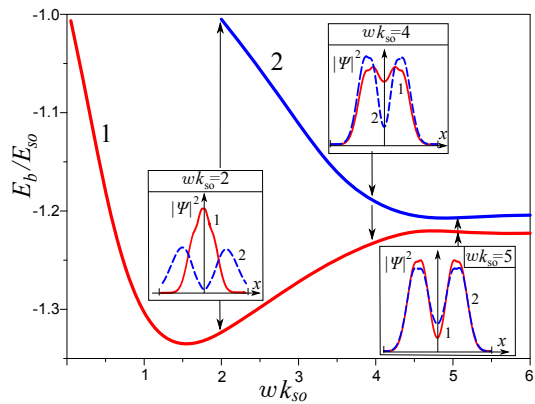}
\caption{(Color online). Energy of the edge-state subband bottoms $E_b$ as functions of the strip width $w$ in the DSOI/RSOI/DSOI structure with the flat potential landscape, $U_D^{left}=U_R=U_D^{right}$, $\alpha=\beta$, $m_R=m_D$. In the insets, the electron density distribution in the edge states is shown for different widths of the strip. Lines 1 and 2 (dashed) correspond to the lower and upper subbands.}
\label{DRDwidth}
\end{figure}

The main difference of the edge-state spectra in the strip structures, as compared with the spectrum in the single RSOI/DSOI contact, is the splitting of the edge-state band into two subbands with different distributions of the electron density across the strip. This is demonstrated in Fig.~\ref{DRDwidth} where the energies $E_{b1}$ and $E_{b2}$ of the edge-state band bottoms in the RSOI/DSOI/RSOI structure are drawn as functions of the strip width $w$. 

It is interesting that the quantization energy depends on the width in unusual manner. The lower subband bottom $E_{b1}$ decreases with increasing $w$ until $wk_{so}\lesssim 1.5$. This is trivially explained by the decrease of the kinetic energy. However, $E_{b1}$ unexpectedly grows as $wk_{so}>1.5$. 

To interpret such a behavior of the quantized energy let us take into consideration the fact that the electron density is redistributed across the strip with increasing $w$. The electron density in the lower subband is redistributed from the center to the edges. As it has been discussed above, electrons gain the energy near the interface because of the mutual action of the SOI in both regions. In the case of the strip, there is an additional energy gain caused by the joint effect of two interfaces. This energy gain decreases with $w$ because the edge states overlap less. It is that reason due to which $E_{b1}$ grows with $w$. In the case of the upper subband this effect is much smaller since the electron density in these states is always concentrated closer to the edges.

It is worth noting that this effect results in an essential lowering of the edge state energy down to the forbidden gap as compared with the case of a single interface.

Above we considered the structures with step-like change in the SOI parameters at the boundaries. So the question arises whether the obtained results persist in the case where the SOI strength changes smoothly in the transition layer between the regions with different SOIs. We studied the problem and found that the edge states survive if the transition layer width $L$ is small, $Lk_{so}<1$. With increasing the layer width the energy interval, where the edge states are located, diminishes to zero and the states disappear. In this point, these edge states differ from the edge states in 2D topological insulators.

This is not only difference between these systems. In contrast to the topological edge states, the states studied here exist in 2D electron gas with one-band spectrum and appear in a finite range of the parameters (the SOI strength, potential step hight, the interface structure parameters). In addition, their spectrum has two pair of branches due to which the interbranch scattering is possible.

\acknowledgments
This work was supported by Russian Foundation for Basic Research and Russian Academy of Sciences (programs ``Quantum Nanostructures''and ``Strongly Correlated Electrons in Semiconductors, Metals, Superconductors, and Magnetic Materials'').


\begin{thebibliography}{20}
\bibitem{note1}
It is known that there is another mechanism of the edge spin currents in 2D systems with heterogeneous SOI. It is caused by the spin-dependent scattering of bulk electrons on the interface.~\cite{Sablikov1} These currents also exist under the equlibrium. The estimations show that the edge-state spin current is small compared to scattering spin current if the Fermi energy lies higher than $E_{so}$ above the conduction band bottom. In opposed case, the edge-state spin current is prevailing, especially when the Fermi level lies in the forbidden band.
\end{thebibliography}
\end{document}